\documentstyle[epsfig,longtable,axodraw]{aipproc}

\newcommand{\dfrac}[2]{\frac{\displaystyle #1}{\displaystyle #2}}

\begin{document}
\title{Universal Torsion Induced Interaction from Large Extra
Dimensions}

\author{Tatsu Takeuchi$\,^a$, Lay Nam Chang$\,^a$, Oleg
Lebedev$\,^a$,\\and\\
Will Loinaz$\,^{a,b}$}
\address{%
$^a$IPPAP, Physics Department, Virginia Tech, Blacksburg, VA 24061\\
$^b$Department of Physics, Amherst College, Amherst, MA 02001%
}

\maketitle

\begin{abstract}
We consider a model with extra compact dimensions in which
only gravity exists in the $4+n$ dimensional bulk.  
If the gravitational connection is not assumed to be
symmetric, that is, unless torsion is artificially set to zero,
then a universal contact interaction among the fermions on the 
4--dimensional wall is induced.
Using a global fit to $Z$--pole observables, we find the
3$\sigma$ bound on the scale of quantum gravity to be
$M_S = 28$~TeV for $n=2$.
If Dirac or light sterile neutrinos are present,
the data from SN1987A increase the bound to
$\sqrt{n} M_S \ge 210$~TeV.
\end{abstract}

Most theories of gravity incorporate the Poincar\'e group as a local symmetry. 
Some, like general relativity, achieve this by demanding local diffeomorphism
invariance, with the additional requirement that gravity can be eliminated
completely around a point by a coordinate choice.  This strong form of the
equivalence principle also restricts the resultant connection coefficient to be
symmetric, thereby eliminating any reference to torsion in the consequent
geometry.  In effect, the theory ``gauges'' the translation group, with the
energy-momentum tensor being the sole source of gravity.  

The more general situation in which the Lorentz group is gauged as well will
give rise to an antisymmetric part to the connection coefficient, the torsion
tensor.  This general
situation obtains in the presence of intrinsic spin; the torsion tensor is then
coupled to the intrinsic spin current, which then represents yet another source
of gravity.  Since this current cannot be eliminated
by a choice of coordinates, the situation breaks the strong form of the
equivalence principle.  It is possible to require that the
torsion tensor still vanishes, but this demand will need to be preserved under
radiative corrections by invoking additional symmetries.  Since the classic
tests of general relativity do not involve intrinsic spin, the presence of
torsion will not show up in any of these checks\cite{wheeler}.

In this paper, we consider the consequences of torsion in the
context of a model with large extra dimensions
\cite{Arkani-Hamed:1998rs}.
Gravity is allowed to exist in a $4+n$ dimensional bulk while all
other fields are confined to a 4--dimensional wall.
Allowing the torsion tensor
$T^{\alpha}{}_{\beta\gamma} = 
\tilde{\Gamma}^{\alpha}{}_{\beta\gamma}-
\tilde{\Gamma}^{\alpha}{}_{\gamma\beta}$ to be non--zero introduces
an extra piece into the gravitational connection
\begin{equation}
\tilde{\Gamma}^{\alpha}{}_{\beta\gamma}=
\Gamma^{\alpha}{}_{\beta\gamma} + K^{\alpha}{}_{\beta\gamma},
\end{equation}
where $\Gamma^{\alpha}{}_{\beta\gamma}$ is the usual metric
contribution,
and $K_{\alpha\beta\gamma} = \frac{1}{2}\left( T_{\alpha\beta\gamma}
-T_{\beta\alpha\gamma}-T_{\gamma\alpha\beta} \right)$ is known
as the contorsion tensor.
The action of our model is given by
\begin{eqnarray}
\lefteqn{
S = -\frac{1}{\hat{\kappa}^2} \int d^{4+n}x \;
     \sqrt{\vert \hat{g}_{4+n}\vert} \;\tilde{R}
} & &  \label{action}
\\
& + & \int d^{4}x \;\sqrt{\vert \hat{g}_4 \vert}\; \frac{i}{2}
         \left[ \bar\Psi\gamma^\mu\tilde\nabla_\mu\Psi
              - \left( \tilde\nabla_\mu\bar\Psi \right)\gamma^\mu\Psi
              + 2 i M \bar\Psi \Psi
         \right].
\nonumber	
\end{eqnarray}	
Here $\hat{\kappa}^2=16\pi G_N^{(4+n)}$, $\tilde{R}$ is the $4+n$
dimensional scalar curvature, and $\hat{g}_{4+n}$ and $\hat{g}_4$
are respectively the $4+n$ and $4$--dimensional (induced) metric
determinants.  The fermion $\Psi$ is coupled to gravity via the
covariant derivative 
$\tilde\nabla_{\mu}\Psi=\partial_{\mu}\Psi
+\frac{i}{2}\tilde\omega^{ab}_{\mu}\sigma_{ab}\Psi$,
where $\tilde\omega^{ab}_{\mu}$ is the spin--connection,
$\sigma_{ab}=\frac{i}{2}\left[ \gamma_a, \gamma_b \right]$, 
with $a$, $b$  the local Lorentz indices. 
A general spin connection $\tilde\omega^{ab}_{\mu}$ can be expressed
in terms  of a torsion-free spin-connection
$\omega^{ab}_{\mu}$, the contorsion tensor, and the vierbein $e^a_\mu$:
\begin{equation}
\tilde\omega^{ab}_{\mu} = \omega^{ab}_{\mu}
+ \frac{1}{4} K^{\nu}{}_{\lambda\mu}
  \left( e^{\lambda a} e^b_\nu - e^{\lambda b} e^a_\nu \right).
\end{equation}
Elimination of torsion from the action by imposing the equations
of motion results in \cite{Chang:2000yw}:
\begin{eqnarray}
S & = & -\frac{1}{\hat{\kappa}^2} \int d^{4+n}x
\;\sqrt{\vert \hat{g}_{4+n} \vert}\;R
        \\  \label{action2}
  &   & + \int d^{4}x \;\sqrt{\vert \hat{g}_4 \vert}\;
        \left[ \bar\Psi\left( i\gamma^{\mu}\nabla_{\mu} - M \right)\Psi
             + \frac{3}{32} 
               \frac{ \sqrt{ \vert \hat{g}_4 \vert } }{ \sqrt{\vert
\hat{g}_{4+n} \vert} } 
               \hat{\kappa}^2\,
               \left( \bar\Psi\gamma_{\mu}\gamma_5\Psi \right)^2\,
               \delta^{(n)}(0) 
        \right]\;. \nonumber
\end{eqnarray}
where $R$ is the torsion-free curvature.
The delta--function appearing in this expression should be regularized
to account for a finite wall thickness:
\begin{eqnarray}
\delta^{(n)}(0) \rightarrow \frac{1}{(2\pi)^n} \int_{0}^{M_S} d^n k =
\dfrac{ M_S^n }{ 2^{n-1} \pi^{n/2}\,n\,\Gamma\!\left(\dfrac{n}{2}\right)
}\;,
\end{eqnarray}
$M_S$ is the cutoff scale of the effective theory, here taken
to be of the order of the inverse wall thinkness. The $4+n$
dimensional coupling constant $\hat{\kappa}$ is related to the
4--dimensional
coupling $\kappa$ and the volume of the extra dimensions compactified on
a torus via $\hat{\kappa}^2=
\kappa^2 V_n=16\pi(4\pi)^{n/2}\Gamma(n/2)M_S^{-(n+2)}$
\cite{Han:1999sg}\footnote{For simplicity we set the string scale
and the $4+n$ dimensional Planck mass equal.}.
As a result, the
leading ${\cal O}\left( \hat{\kappa}^2 \right)$ torsion contribution
to the action is given by
\begin{equation}
\Delta S
= \int d^4x\;\frac{3\pi}{n M_S^2}
      \left[ \sum_j \bar\Psi_j\gamma_{\mu}\gamma_5\Psi_j
      \right]^2\;,
\label{action3}
\end{equation}
where $j$ runs over all fermions existing on the wall.
The expansion in $\hat{\kappa}$ is expected to be valid provided the
typical energy $E$
of a physical process is  below the cutoff scale $M_S$.

\begin{center}
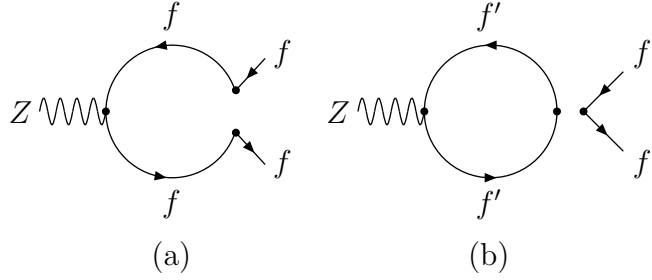
\begin{figure}[t]
\begin{center} 
\begin{picture}(350,110)(-40,-20)
\Vertex(50,50){1.5}
\Vertex(99,58){1.5}
\Vertex(99,42){1.5}
\Photon(25,50)(50,50){5}{4}
\ArrowArc(75,50)(25,20,180)
\ArrowArc(75,50)(25,180,340)
\ArrowLine(110,70)(99,58)
\ArrowLine(99,42)(110,30)
\Text(18,50)[]{$Z$}
\Text(75,86)[]{$f$}
\Text(75,15)[]{$f$}
\Text(117,72)[]{$f$}
\Text(117,30)[]{$f$}
\Vertex(170,50){1.5}
\Vertex(220,50){1.5}
\Vertex(230,50){1.5}
\Photon(145,50)(170,50){5}{4}
\ArrowArc(195,50)(25,0,180)
\ArrowArc(195,50)(25,180,360)
\ArrowLine(245,65)(230,50)
\ArrowLine(230,50)(245,35)
\Text(138,50)[]{$Z$}
\Text(195,86)[]{$f'$}
\Text(195,15)[]{$f'$}
\Text(253,72)[]{$f$}
\Text(253,30)[]{$f$}
\Text(75,-5)[]{(a)}
\Text(195,-5)[]{(b)}
\end{picture}
\end{center}
\caption{Corrections to the $Zf\bar{f}$ coupling from the universal 
contact interaction Eq.~(\ref{action3}).} 
\label{Feynman}
\end{figure}
\end{center}

The torsion induced contact interaction Eq.~(\ref{action3}) can
be constrained through its effect on
$Z$--pole electroweak observables.
The corrections shown in Fig.~1 shifts the $Z$--couplings by
\begin{equation}
\delta h_L=-\delta h_R={3N_c m_t^2 \over 4\pi n M_S^2}\;
\label{delta}
\ln {M_S^2 \over m_t^2}\;,
\end{equation}
where the Z--vertex is defined as $-i{g\over \cos \theta_W}
Z_\mu \bar\Psi \gamma^\mu (P_Lh_L+P_Rh_R)\Psi$.  Note that the 
contribution of torsion to $\delta h_L$ is strictly positive.
Performing a global fit to the LEP/SLD electroweak observables
will lead to a constraint on $\delta h_L$, which in turn will
give us a limit on $M_S$.
For $n=2$ we find the $3\sigma$ bound to be
\begin{equation}
M_S \geq 28 \; {\rm TeV}.
\end{equation}
For $n=4 (6)$ the bound weakens to 19 (15) TeV.
The details of the analysis are presented 
elsewhere \cite{Chang:2000yw}.

Other constraints are available as well. The universal
$A\times A^{+}$ contact interaction Eq.~(\ref{action3}) affects
at the tree level
the differential cross sections for $e^+e^- \rightarrow f\bar f$
measured at LEP.
The OPAL measurements \cite{opal} imply
\begin{equation}
\sqrt{n} M_S \geq 10.3\;{\rm TeV}
\end{equation}
at the 95\% confidence level. Electron--quark contact interactions can
also be constrained via HERA DIS data, Drell--Yan production at the
Tevatron, {\it etc.}
The global analysis \cite{cheung} yields
\begin{equation}
\sqrt{n} M_S \geq 5.3\;{\rm TeV}\;.
\end{equation}
Another potentially strong constraint can come from the measurement
of the invisible width of the $\Upsilon$ and $J/\Psi$ resonances
at B and $\tau$--charm factories \cite{Chang:1998tq}.

A powerful astrophysical constraint can be derived if we admit 
existence of Dirac or light sterile neutrinos.
For the case of Dirac neutrinos, the torsion--induced interaction
containes a term
\begin{equation}
\Delta {\cal{L}}=
-\frac{6\pi}{n M_S^2}\;
\bar q\gamma^{\mu}\gamma_5 q \; \bar \nu_R\gamma_{\mu} \nu_R\;.
\end{equation}
This contact interaction provides a new channel of energy drain
during neutron star collapse, since right handed neutrinos
produced by nucleon interactions leave the core without rescattering
This would affect neutron star
evolution; in particular, it would modify the duration of the standard
neutrino burst. From observations of SN 1987A one infers \cite{Grifols:1998iy}
\begin{equation}
\sqrt{n} M_S \geq 210\;{\rm TeV}\;.
\label{SN}
\end{equation}
Similar considerations apply to the case of light sterile neutrinos:
if $m_{\nu_s} \ll 50\;{\rm MeV}$, the core temperature, the analysis is
completely analogous to that of Dirac neutrinos and the bound
(\ref{SN}) holds.  This bound translates into an upper bound 
on the compactification radius of $3 \times 10^{-5}$ mm for $n=2$.

Since $M_S$ controls all gravity effects in extra dimensions, the limits
on $M_S$ being larger than tens of TeV
reported here imply weaker KK graviton couplings than those considered
in
the literature.  


This research is supported in part by a grant from 
the U.S. Department of Energy, DE--FG05--92ER40709.



\begin{references}


\bibitem{wheeler} I.~Ciufolini, J.~A.~Wheeler, 
{\it Gravitation and Inertia} (Princeton University, New Jersey,
1995).

\bibitem{Arkani-Hamed:1998rs}
N.~Arkani--Hamed, S.~Dimopoulos and G.~Dvali,
Phys.\ Lett.\  {\bf B429}, 263 (1998); Phys.\ Rev.\  {\bf D59}, 086004
(1999);
I.~Antoniadis, N.~Arkani--Hamed, S.~Dimopoulos and G.~Dvali,
Phys.\ Lett.\  {\bf B436}, 257 (1998).

\bibitem{Chang:2000yw}
L.~N.~Chang, O.~Lebedev, W.~Loinaz and T.~Takeuchi,
Phys.\ Rev.\ Lett.\  {\bf 85}, 3765 (2000)
[hep-ph/0005236].

\bibitem{Han:1999sg}
T.~Han, J.~D.~Lykken and R.~Zhang,
Phys.\ Rev.\  {\bf D59}, 105006 (1999).


\bibitem{opal}
OPAL Collaboration, G.~Alexander {\it et al.},
Phys. Lett. {\bf B387}, 432 (1996); G.~Abbiendi {\it et al.},
Eur.\ Phys.\ J.\  {\bf C6}, 1 (1999).

\bibitem{cheung} 
K.~Cheung, hep-ph/9807483; 
V.~Barger, K.~Cheung, K.~Hagiwara,
D.~Zeppenfeld, Phys. Rev. {\bf D57}, 391 (1998).

\bibitem{Chang:1998tq}
L.~N.~Chang, O.~Lebedev and J.~N.~Ng,
Phys.\ Lett.\  {\bf B441}, 419 (1998)
[hep-ph/9806487].

\bibitem{Grifols:1998iy}
J.~A.~Grifols, E.~Mass\'{o} and R.~Toldr\`{a},
Phys.\ Rev.\  {\bf D57}, 2005 (1998)
[hep-ph/9707531].


\end{references}
\end{document}